\documentclass[prb,aps,twocolumn,superscriptaddress,longbibliography]{revtex4-2}

\usepackage{amsmath,amssymb}
\usepackage{graphicx}
\usepackage{bm}
\usepackage{hyperref}

\graphicspath{{figures/}}
\begin{document}

\title{Even/Odd-parity STS spectra induced by quantum well mirror symmetry breaking in iron-based superconductors}

\author{Xiuqing Huang}
\email{xiuqing\_huang@163.com}

\affiliation{Department of Telecommunications Engineering ICE, Army Engineering University of PLA, Nanjing 210007, China}
\affiliation{National Laboratory of Solid State Microstructures of Physics, Nanjing
	University, Nanjing 210093, China}

\date{\today}

\begin{abstract}
		
	Determining whether superconducting scanning tunneling spectroscopy (STS) is uniquely dictated by crystal structure constitutes a fundamental challenge in condensed matter physics. Here, we systematically investigate bulk FeSe single crystals, monolayer FeSe, and $\mathrm{KCa_2Fe_4As_4F_2}$. We resolve one-, two-, and three-order checkerboard quantum-well structures that perfectly match the experimentally observed one, two, and three pairs of superconducting coherence peaks. In bulk FeSe, quantum wells promote real-space Cooper pairing and form degenerate antiferromagnetic checkerboard sublattices, yielding bosonic even-parity STS responses. In monolayer FeSe, mirror symmetry breaking suppresses Cooper pairing and induces nondegenerate ferromagnetic sublattice dichotomy, producing fermionic odd-parity STS spectra. We establish a universal gap scaling law $\Delta(T, \xi) = \eta(T)/\xi^2$, where $\eta(T)$ is a temperature-dependent prefactor and $\xi$ denotes quantum-well depth that governs the number and magnitude of superconducting gaps. For $\mathrm{KCa_2Fe_4As_4F_2}$, our predicted gap pairs of $\pm6.2$ meV, $\pm5.6$ meV, and $\pm4.2$ meV are in excellent agreement with experimental results of $\pm6.2$ meV, $\pm5.4$ meV, and $\pm4.4$ meV. This quantum-well mechanism unifies mirror symmetry breaking, checkerboard sublattice ordering, Cooper pairing, fermion-boson duality, and half-Bogoliubov states for STS interpretation, offering new insights toward a unified high-$T_\text{c}$ superconductivity theory.

\end{abstract}

\maketitle

\section{Introduction}

From cuprate, iron-based to nickelate superconductors, a variety of unconventional superconducting systems have been discovered \cite{bednorz1986possible,wu1987superconductivity,schilling1993superconductivity,kamihara2008iron,takahashi2008superconductivity,ren2008superconductivity,hsu2008superconductivity,anisimov1999electronic,li2019superconductivity,sun2023superconductivity,Zhang2025Nickelate96K}. Despite substantial advances in experimental detection techniques \cite{Keimer2015Nature,Davis2011NJP}, a self-consistent physical picture for high-$T_\text{c}$ superconductivity remains absent \cite{Anderson2007Science}, and its intrinsic nature and microscopic pairing mechanism constitute an unresolved core problem in condensed matter physics \cite{Zaanen2006NatPhys,Dagotto2013RMP}. 
Scanning tunneling microscopy/spectroscopy (STM/STS) is a powerful local probe for unconventional superconductors, having enabled landmark microscopic discoveries over the past four decades \cite{Fischer2007RMP}, including pseudogap states \cite{Pan2001Nature}, charge-density wave order \cite{Howald2003PNAS}, electronic nematic order \cite{Kohsaka2007ScienceNematicSTM}, integer and fractional flux vortices \cite{Hess1989PRL,Zheng2026vortexfrac}, intertwined spin-charge stripe domains \cite{Hanaguri2004NatureStripeSTM}, half-Bogoliubov quasiparticle states \cite{Li2026HalfBogo}, and nanoscale electronic phase separation \cite{Lang2002PRL}. 
Nevertheless, a key research bottleneck persists: no theoretical model can establish a quantitative correspondence between nanoscale tunneling spectra and macroscopic superconducting transport properties \cite{Bozovic2016Nature}. In particular, it remains unclear how periodic lattice and charge modulations jointly regulate the position, spectral weight, number and symmetry of quasiparticle interference peaks in STS conductance spectra.

Moreover, prevailing high-$T_\text{c}$ theoretical frameworks adopt drastic simplifications. Representative minimal low-energy models are the $t$–$J$ model \cite{OgataFukuyama2008tJ,Dagotto1994RevModPhys} and Hubbard model \cite{Hubbard1963ProcRSocA,Lieb1989TwoTheoremsHubbard,Qin2022HubbardComp}, together with broad phenomenological quasiparticle scattering theories \cite{Kivelson2023RevModPhys,Hoffman2002ImagingQPI}. Such idealized models cannot fully capture strong-correlation-governed real-space electronic responses. While BCS-derived momentum-space band analysis is routinely used to interpret tunneling and spectroscopy data \cite{Bardeen1957PhysRev}, it fails to uncover real-space dynamic electron interactions driving high-$T_\text{c}$ pairing \cite{Shen2003RevModPhys}. 
Mounting experimental evidence proves emergent behaviors of strongly correlated systems arise from real-space localized electrons \cite{Hoffman2002VortexChecker4UC,Ghiringhelli2012LongRangeCDWfluct,Zou2024ChargeOrderInsulatorSC,Wise2008CheckerCDWOrigin,Wan2021FeIncommAFM}. The community widely agrees that any consistent high-$T_\text{c}$ theory must incorporate a Mott insulator framework accounting for strong electronic localization \cite{Anderson1987ScienceRVB,Davis2022Science}. High-$T_\text{c}$ superconductivity emerges via doping Mott parent compounds, which unavoidably generates quenched disorder. The favorable correlation between disorder and superconducting strength discredits ideal-Fermi-surface free-electron band theories for unconventional superconductors. Complementary STS studies on iron-pnictide and nickelate systems further identify real-space electronic localization as a universal hallmark of all strongly correlated unconventional superconductors \cite{Fernandes2022Nature,Wang2025NSR}.

Recently, we proposed a quantum-well localized electron superconductivity theory based on the lossless Planck quantum ground state \cite{huang2024rs}, deriving a universal optimal $T_\text{c}$ formula for unconventional superconductors \cite{huang2025arxiv}:
\begin{equation}
	T_\text{c} = \frac{\Lambda}{\xi^2}, \label{eq:Tc}
\end{equation}
where $\xi$ is the quantum well depth, $\Lambda(\text{Cu}) \simeq 1300$, and $\Lambda(\text{Fe}) \simeq 400$. 
With $\xi$ values determined by X-ray measurements, we accurately predict the optimal $T_\text{c}$ for cuprate, iron-based, and nickel-based superconductors, in good agreement with experimental observations.

This work explores the relation between $dI/dV$ spectral symmetry and quantum-well mirror symmetry breaking in iron-based superconductors using quantum-well-localized superconducting electrons.
Cuprate and nickel-based superconductors only have Cu-O/Ni-O superconducting planes, while iron-based materials possess a three-layer folded structure (upper/lower As(Se) + middle Fe) with three unique quantum-well configurations and superconducting states.
Without the two-Fe sublattice hypothesis\cite{Hu2013,Hirschfeld2011,Ding2026PRL,Xu2026CPL,Roig2025OriginSPHA}, we demonstrate that bulk FeSe\cite{Kreisel2020,Wang2021}, monolayer FeSe\cite{Xue2012,Ding2024}, and $\text{KCa}_2\text{Fe}_4\text{As}_4\text{F}_2$\cite{Wang2016JACS,Duan2021PRB} host one-, two-, three-layer checkerboard sublattices corresponding to three sets of superconducting STS gaps. Bulk FeSe’s quantum wells with Cooper pairs produce bosonic antiferromagnetic sublattice dichotomy and even-parity STS signals. In contrast, monolayer FeSe on SrTiO$_3$ experiences interface-induced mirror symmetry breaking, which disrupts real-space Cooper pairing and antiferromagnetic order to form odd-parity STS spectra.
Combined with Eq. (\ref{eq:Tc}) and the linear relation between gap $2\Delta$ and $T_\text{c}$, our model reproduces both qualitative and quantitative STS features of the three compounds. The fitted position, count and parity of superconducting interference peaks agree well with experimental data. This quantum-well framework accurately predicts $T_\text{c}$ and superconducting gaps, and links spectral parity, real-space Cooper pairs, fermion-boson crossover, half-Bogoliubov states and quantum-well geometry. It offers a new perspective for unconventional superconductivity and advances a unified high-$T_\text{c}$ theoretical framework.

\section{Quantum Well Symmetry Breaking and STM Spectroscopy}

\begin{figure}[tpb]
	\centering
	\includegraphics[width=\columnwidth]{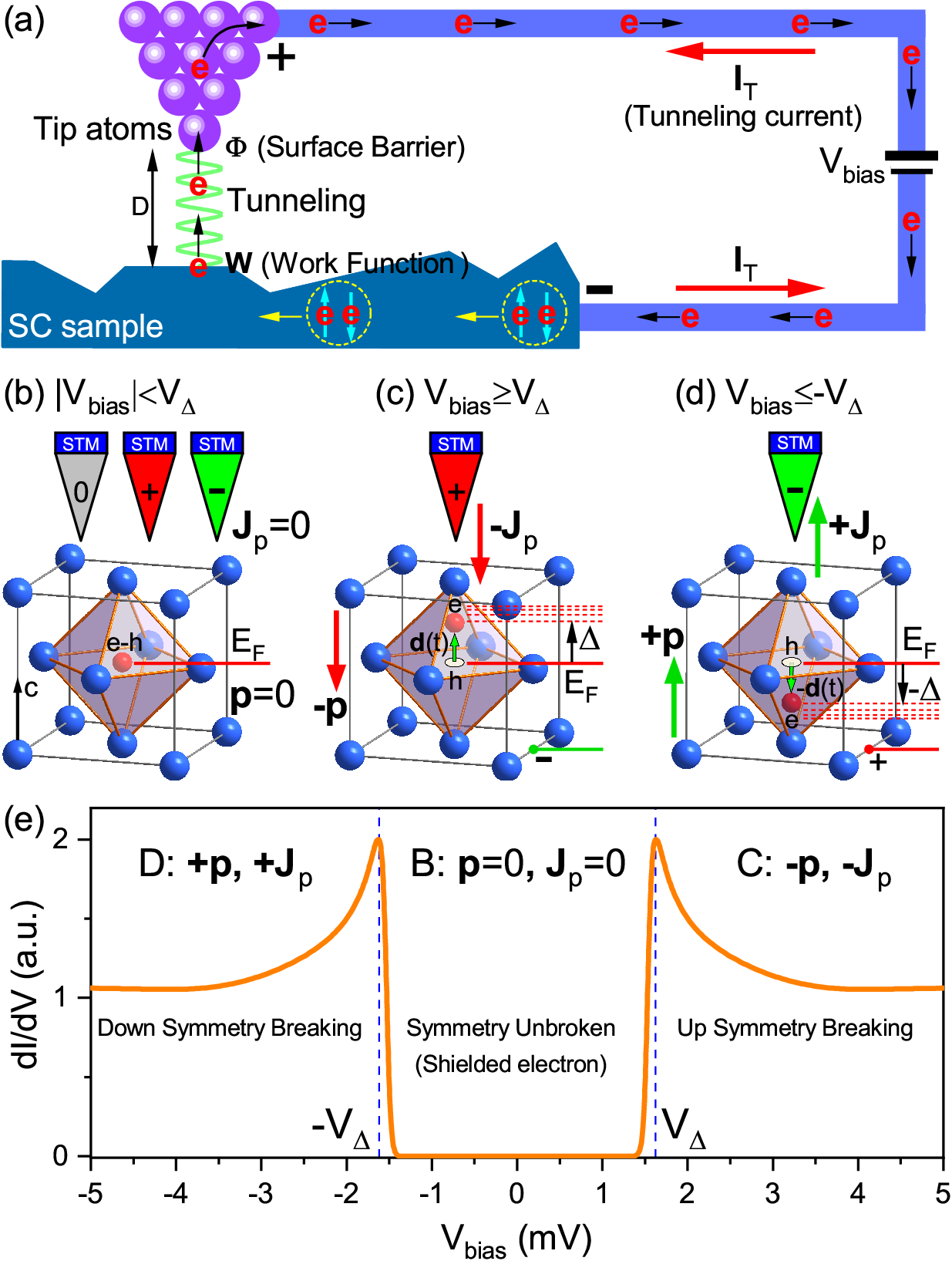}
	\caption{
		Physical nature of STM spectroscopy: electron versus electromagnetic wave tunneling.
		(a) Conventional free-electron theory attributes STM tunneling to energy-conserving wavefunction-mediated electron transport. However, $eV_{\text{bias}} \ll W,\Phi$, which is insufficient for electrons to traverse the vacuum gap $D$.
		(b)--(d) This work demonstrates that STM current originates from bias-induced polarization rather than intrinsic electron tunneling.
		(b) At zero/weak bias, electrons stay below the threshold $\Delta$, localizing at $E_\text{F}$ without dipole formation or polarization current.
		(c) Positive bias above $\Delta$ breaks electronic symmetry, inducing upward electron displacement $d(t)$ and dipoles $-\boldsymbol{p}$ to produce downward current $-J_p$.
		(d) Negative bias induces downward displacement $-d(t)$ and reversed dipoles $+\boldsymbol{p}$, yielding upward current $+J_p$.
		(e) The simulated conductance spectra in regions B, C, D correspond to the states in (b), (c), (d).
		Essentially, electrons function as bias-controlled tunneling switches via symmetry breaking rather than tunneling carriers. The actual tunneling originates from the electromagnetic field of the external bias.
	}
	\label{figure1}
\end{figure}

Conventional scanning tunneling microscopy (STM) acquires signals relying on electron quantum tunneling \cite{Fischer2007RMP}. Fig. \ref{figure1}(a) sketches an STM device under positive sample bias. Quantum theory claims millivolt-scale $V_{\text{bias}}$ allows periodic Cooper-pair tunneling across the vacuum gap. We verify this model’s physical validity via energy conservation analysis. Tunneling requires electrons to cross two successive high-energy barriers: the superconductor work function $W$ (eV magnitude) to enter vacuum spacing $D$, plus the surface potential barrier $\Phi$ (tens of eV) en route to the probe tip. Only a millivolt bias field provides external energy, and electron wavefunctions carry no extra energy input. This fundamental mismatch yields a critical doubt: whether the standard tunneling picture contradicts energy conservation?

\subsection{Tunneling: Electrons or Electromagnetic Waves}

As information carriers, electrons and photons (electromagnetic waves) are essential to both scientific research and daily life. Photon tunneling is a ubiquitous phenomenon, existing in natural lightning as well as man-made technologies including Wi-Fi, radar, microwave heating, X-ray imaging and CT scanners. Since Kamerlingh Onnes discovered dissipationless superconducting transport in 1911, more than a century of superconductivity research has revolved around a core puzzle: how electrons achieve lossless tunneling?

Superconductivity refers to a dissipationless condensed phase, and superconducting currents must satisfy continuity, stability, and closed-loop conditions simultaneously. We propose that superconducting charge carriers propagate force-free with no energy exchange with the ambient environment to sustain superconductivity and quantum tunneling; such carriers are herein defined as \textbf{energy insulators}. Accordingly, we derive five mandatory constraints governing carriers that produce tunneling current $I_T$:

\begin{itemize}
	\item Neutral, unaffected by Coulomb forces;
	\item Zero rest mass, gravity-independent;
	\item Point-like geometry without interparticle collisions;
	\item No heat exchange, temperature-invariant propagation speed;
	\item Barrier penetration unlimited by energy-scale constraints.
\end{itemize}

Massless neutral photons, i.e., electromagnetic waves, propagate at the invariant speed of light and comply with all five constraints, whose propagation is independent of temperature and energy thresholds. As a common physical fact, light can transmit through flowing liquid water and remains transmissible within ice after water freezes upon cooling. In contrast, electrons bear electric charge, have finite rest mass and spatial extent, and break all the aforementioned constraints.
Superconductivity is realized by suppressing electron thermal motion at low temperatures. Once entering the superconducting state, lattices and valence electrons condense into a rigid aggregate, and electrons lose the capability of free motion and tunneling. The assumption that electrons contribute tunneling current has no scientific foundation. Accordingly, we propose that STM signals originate from electromagnetic waves instead of free-electron tunneling. 

Fig. \ref{figure1}(b) illustrates octahedral quantum wells confining electrons in fcc lattices, where $\Delta$ denotes the coupling strength between electrons and quantum wells. Charge neutrality requires one lattice hole per quantum well. Under weak bias with $T<T_\text{c}$, electrons undergo symmetric screening and reside steadily at the ground-state Fermi level $E_\text{F}$. 
As illustrated in Fig. \ref{figure1}(c), positive bias drives electrons past threshold $\Delta$ and breaks symmetry, yielding dipole moment $-\boldsymbol{p}=-e\boldsymbol{d}(t)$. Negative bias reverses electron displacement and dipole moment $+\boldsymbol{p}=e\boldsymbol{d}(t)$. 
Under the polarization current framework, tunneling current originates from system polarization instead of long-range free-charge transport, with current density expressed as:

\begin{equation}
	\boldsymbol{J}_P = \boldsymbol{J}_T 
	= \pm n \frac{\partial \boldsymbol{p}}{\partial t} 
	= \pm n e \frac{\partial \boldsymbol{d}(t)}{\partial t},
	\label{eq:IT}
\end{equation}
where $\boldsymbol{p}$ is the dipole moment of a confined electron, $e$ elementary charge, $n$ quantum-well volume density. Positive sign denotes upward dipoles under negative bias; negative sign downward dipoles under positive bias.

Combining Fig.~\ref{figure1} and Eq.~(\ref{eq:IT}), the polarization current is parallel to the dipole polarization of electrons and antiparallel to the applied electric field, with its magnitude linearly proportional to the bias electric field. It is thus demonstrated that tunneling essentially corresponds to bias-driven dynamic local displacement of electrons, independent of static uncoupled wavefunctions.
Fig.~\ref{figure1}(e) displays the qualitative differential conductance spectrum of scanning tunneling microscopy (STM) derived from Fig.~\ref{figure1}(b)--(d), which clearly reveals electric-field-induced symmetry breaking and dipole dynamics of quantum-well-bound electrons. Three electronic regimes are identified: large positive bias (Regime C) shifts electrons upward, negative bias (Regime D) induces downward electron displacement, while Regime B features static localized electrons. The spectral symmetry is entirely governed by the mirror symmetry of the quantum well. This conclusion rules out conventional tunneling models based on n-type electrons and p-type holes. Conventional STM measurements merely capture superconducting coherence peaks, whose underlying microscopic mechanism remains elusive. This work elucidates the dynamical picture of bias-tunable electron displacement and symmetry breaking in spectral signals from confined quantum-well electrons.

Superconducting gap correlates with temperature, doping and pressure, characterizing coupling strength between localized electrons and quantum wells. Lower temperature strengthens electron localization and interwell coupling to enlarge gap magnitude. For quantum wells of depth $\xi$, zero-temperature maximum gap $\Delta_0(\xi)$ scales linearly with $T_\mathrm{c}$. This linear relation is formulated by Eq.~\eqref{eq:Tc}, ignoring pairing symmetry and minor secondary effects.

\begin{equation}
	\Delta(T, \xi) = \frac{\eta (T)}{\xi^2}= \Delta_0(\xi) \tanh\left( \alpha_0 \sqrt{\frac{T_c}{T} - 1} \right),
	\label{eq:delta}
\end{equation}
where  $\eta(T)$ is an undetermined constant associated with material characteristics and temperature, $\alpha_0$ is a dimensionless phenomenological parameter.

\begin{figure}[t]
	\centering
	\includegraphics[width=\columnwidth]{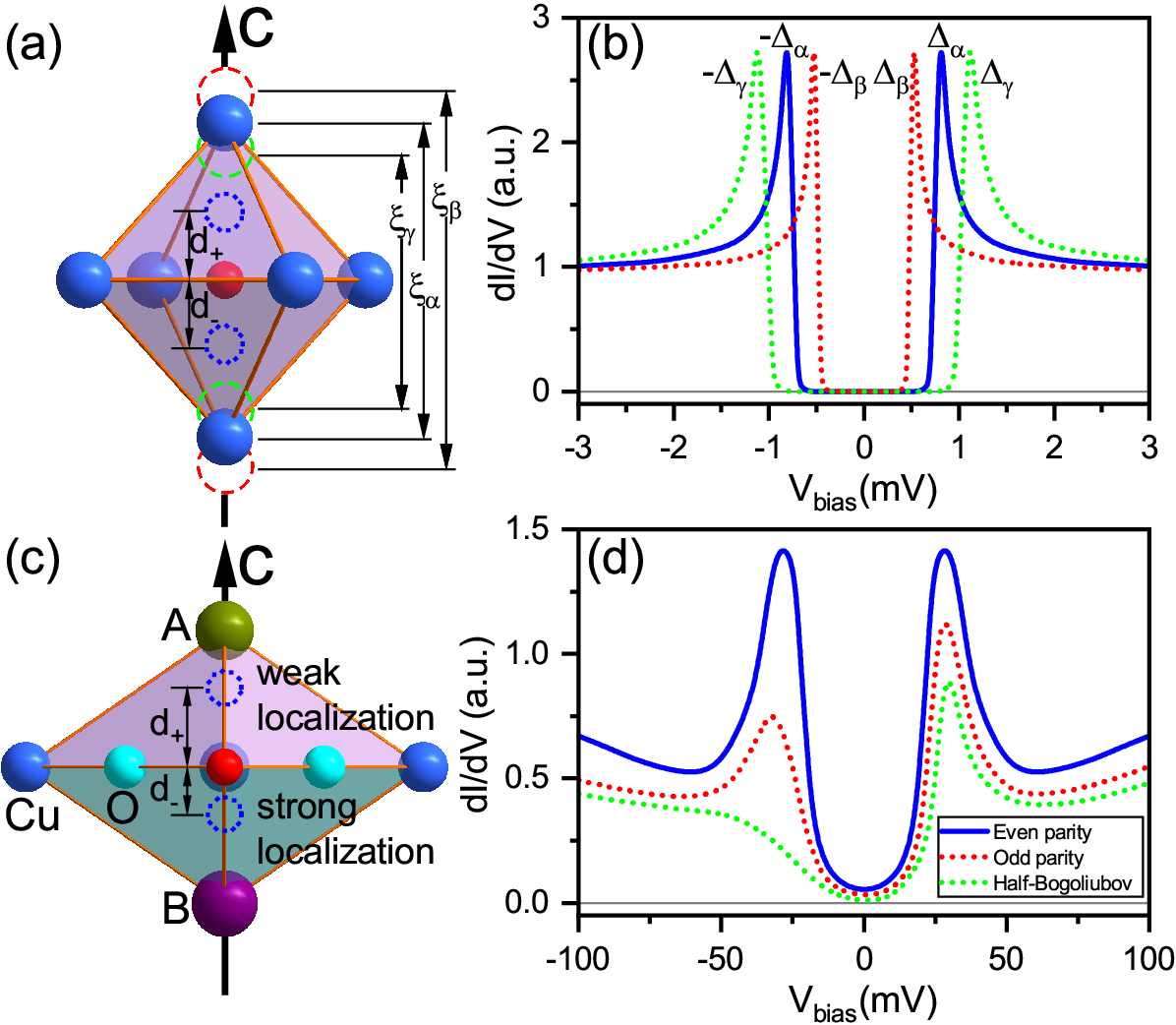}
	\caption{Correlation between quantum well structures and parity symmetry of STS spectra.
		(a,b) Electrons confined in mirror-symmetric quantum wells show identical displacements $d_{+}=d_{-}$ under symmetric biases, producing even-parity symmetric $dI/dV$ spectra. Superconducting coherence peak energy scales inversely with squared well depth, obeying $\xi_\gamma<\xi_\alpha<\xi_\beta$ (well depths) and $\Delta_\beta<\Delta_\alpha<\Delta_\gamma$ (gaps).
		(c,d) Doping breaks mirror symmetry of cuprate quantum wells and forms asymmetric confinement potentials. A weaker upper barrier strengthens upward electron drift ($d_{+}>d_{-}$), generating odd-parity STS spectra with reduced left spectral weight and enhanced right weight (red dashed lines). At ultralow doping, the undoped lower half-well exhibits strong insulating localization, nearly vanishing downward displacement $d_{-}\to 0$ and yielding half-Bogoliubov features. A reinforced upper barrier reverses this spectral asymmetry.}
	\label{figure2}
\end{figure}

\subsection{Mirror Symmetry Breaking and Parity Symmetry}

Fig. \ref{figure2}(a) illustrates mirror-symmetric octahedral quantum wells in conventional superconductors. Without applied bias, Coulomb-screened electrons localize at the well center, yielding an insulating state. Upon bias application, electrons shift along the Coulomb force, breaking the $c$-axis mirror symmetry and generating polarization currents. Positive and negative biases induce symmetric symmetry breaking with equal electron displacements ($d_{+}=d_{-}$) and opposite polarization current directions, yielding symmetric even-parity coherence peaks of identical amplitude $\pm\Delta_\alpha$, as shown by the solid blue curve in Fig. \ref{figure2}(b) \cite{Pan1998a,Cherkez2014Proximity,Ast2016AlSTS}.
According to Eq.~(\ref{eq:delta}), $c$-axis quantum well expansion ($\alpha\to\beta$) narrows the paired $\mathrm{d}I/\mathrm{d}V$ peaks inward, corresponding to the red dashed curve ($\pm\Delta_\beta$) in Fig. \ref{figure2}(b). In contrast, quantum well compression ($\alpha\to\gamma$) broadens the twin peaks outward, shown by the green dashed curve ($\pm\Delta_\gamma$). 
For the unconventional superconductor displayed in Fig. \ref{figure2}(c), doping destroys $c$-axis mirror symmetry and distorts quantum well geometry. Identical apical A and B ions preserve $c$-axis symmetry and give rise to twin symmetric even-parity peaks in the $\mathrm{d}I/\mathrm{d}V$ spectrum (solid blue curve in Fig. \ref{figure2}(d)) \cite{Misra2002a,Hoogenboom2003a,Matsuba2003a}. Doping-induced asymmetry between apical A and B ions breaks mirror symmetry and creates uneven confinement strength between upper and lower quantum wells. If the lower well confines electrons more strongly, downward electron motion is hindered ($d_{+}>d_{-}$), leading to asymmetric odd-parity peaks with suppressed left-side spectral weight and enhanced right-side intensity (red dashed curve). 

In heavily underdoped La-Bi2201 \cite{Li2026HalfBogo}, the lower half of the quantum well retains the Mott insulating phase of the undoped parent compound. Downward electron displacement is almost fully suppressed ($d_{-}\rightarrow 0$), which blocks the formation of downward dipoles and eliminates left coherence peaks, leaving only a unilateral right half-Bogoliubov tunneling peak (green dashed curve). By analogy, weakly doped or undoped upper vertices preserve Mott insulating behavior in the upper quantum well, imposing stronger electron confinement and producing asymmetric spectral peaks with elevated left intensity and reduced right intensity. In the extreme limit, solely a left half-Bogoliubov tunneling peak remains.

\section{Intrinsic Symmetry Breaking in Iron-Based Quantum Wells}

\begin{figure}[t]
	\centering
	\includegraphics[width=\columnwidth]{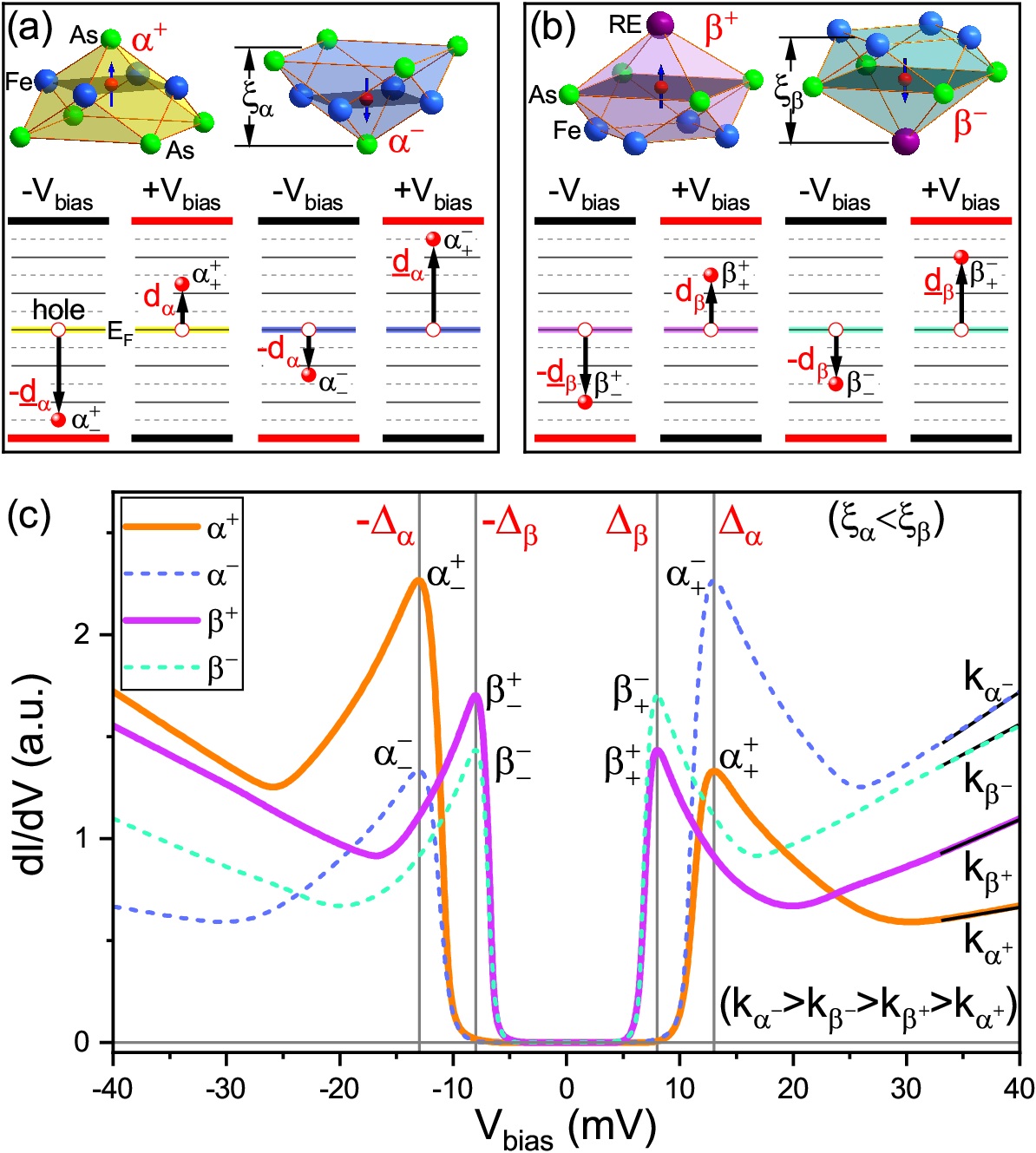}
	\caption{
		Intrinsic mirror symmetry breaking and odd-parity STS spectral symmetry in iron-based superconductor quantum wells
		(a) Tent-shaped octahedral quantum wells host upward \(\alpha^{+}\) and downward \(\alpha^{-}\) excited states. Under bias, Fe-site Fermi-level \(E_\mathrm{F}\) ground-state electrons shift in electric fields, yielding four confinement-tunable electron-hole dipole states \((\alpha^{+}_{\pm},\alpha^{-}_{\pm})\).
		(b) Gem-like tridecahedral quantum wells contain upward \(\beta^{+}\) and downward \(\beta^{-}\) states. Bias-induced Coulomb forces displace As-layer electrons to generate four confinement-tunable dipole states \((\beta^{+}_{\pm},\beta^{-}_{\pm})\).
		(c) Well depths \(\xi_\alpha,\xi_\beta\), geometry, orientation and dipole strength determine resultant \(dI/dV\) spectra: solid orange for \(\alpha^{+}\), dashed blue for \(\alpha^{-}\), solid purple for \(\beta^{+}\), dashed green for \(\beta^{-}\). Individual spectra exhibit asymmetric odd parity, while paired counteroriented wells (\(\alpha^{+}+\alpha^{-}\), \(\beta^{+}+\beta^{-}\)) compose symmetric even-parity composite STS spectra.}
	\label{figure3}
\end{figure}

The uniqueness of iron-based superconductors stems from their characteristic folded triple-layer As(Se)/Fe/As(Se) structure \cite{kamihara2008iron}. In contrast to the octahedral quantum wells in cuprate superconductors (Fig. \ref{figure2}(c)), iron-based hosts two distinct quantum well geometries: tent-shaped nonahedral wells with Fe layers as superconducting planes (Fig. \ref{figure3}(a)), and gem-like tridecahedral wells where superconductivity resides on As layers (Fig. \ref{figure3}(b)).
Devoid of intrinsic mirror symmetry, such structures host localized electronic ground states featuring positive/negative energy levels $\alpha^\pm$ and $\beta^\pm$ with reversed orientations, which correspond to spin-up and spin-down electrons, respectively. Given $\xi_\alpha < \xi_\beta$, iron-based superconductors generally exhibit two distinct superconducting states with different energy gaps and critical temperatures. According to Eq.~\eqref{eq:delta}, it can be readily concluded that $\Delta_\alpha > \Delta_\beta$ and $T_\mathrm{c}^\alpha > T_\mathrm{c}^\beta$.
Unlike cuprate octahedral quantum wells with two apical sites, iron-based quantum wells confine electrons via one apical site and one square face, inducing asymmetric electron displacement polarization. As displayed in the lower panels of Fig. \ref{figure3}(a) and (b), symmetry breaking generates eight electric dipoles corresponding to four pairs of excited states: $(\alpha^{+}_{-},\alpha^{-}_{+})$, $(\alpha^{+}_{+},\alpha^{-}_{-})$, $(\beta^{+}_{-},\beta^{-}_{+})$, and $(\beta^{+}_{+},\beta^{-}_{-})$. These states exactly match the four pairs  of superconducting coherence peaks in Fig. \ref{figure3}(c). 

Intrinsic structural asymmetry between the upper and lower mirror surfaces of iron-based superconducting quantum wells induces anisotropic quantum confinement, which is fully reflected in corresponding $dI/dV$ spectra. Take the upward-pointing yellow $\alpha$-type quantum well in Fig. \ref{figure3}(a) as an example: localized electrons are tightly confined by apical arsenic atoms, yielding minimal upward displacement $d_\alpha$ under bias and the lowest $\alpha^{+}_{+}$ peak in Fig. \ref{figure3}(c). In contrast, four basal As atoms provide weak long-range indirect confinement, generating maximum displacement $\underline{d}_\alpha$ and the dominant $\alpha^{+}_{-}$ peak. 
For the purple $\beta$-type quantum well in Fig. \ref{figure3}(b), apical rare-earth ions sit farther from electrons than apical As in $\alpha$ wells, weakening confinement and giving $d_\beta > d_\alpha$, so the $\beta^{+}_{+}$ peak exceeds $\alpha^{+}_{+}$. Meanwhile, basal Fe atoms lie closer to electrons than basal As in $\alpha$ wells, strengthening confinement such that $|-\underline{d}_\beta| < |-\underline{d}_\alpha|$, leading to a weaker $\beta^{+}_{-}$ peak relative to $\alpha^{+}_{-}$.  
At high bias, electron displacement scales linearly with applied electric field, and differential conductance rises linearly with bias. Stronger confinement suppresses displacement variation, reducing $dI/dV$ growth and yielding gentler spectral slopes. The high-bias slopes follow $k_{\alpha^{-}} > k_{\beta^{-}} > k_{\beta^{+}} > k_{\alpha^{+}}$. 
In summary, STM spectral features are analytically derived solely from quantum-well geometric properties. Subsequent sections verify these theoretical predictions using three representative iron-based superconductors.

\section{Even Parity and Cooper-Pair Sublattice Dichotomy in Single-Crystal FeSe}

\begin{figure}[t]
	\centering
	\includegraphics[width=\columnwidth]{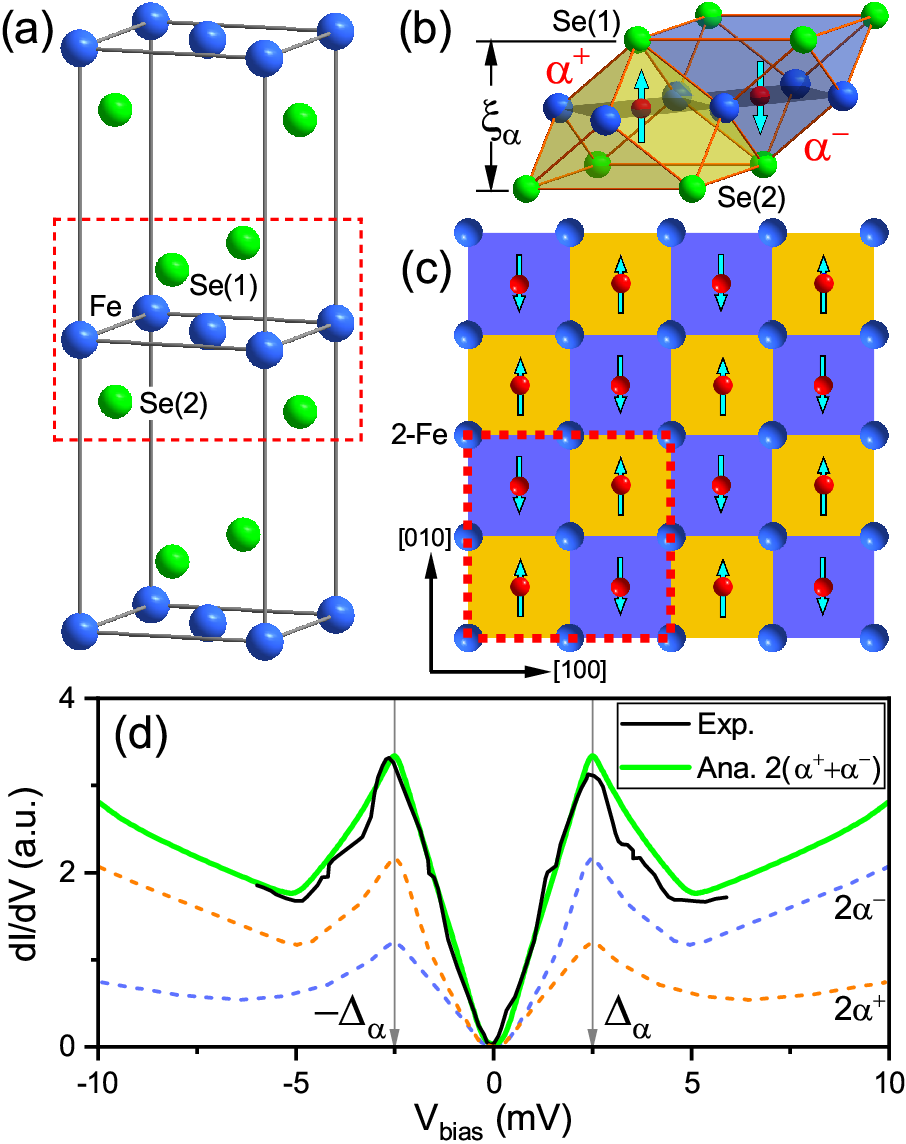}
	\caption{
		Quantum well sublattice dichotomy and even-parity STS spectra in FeSe single crystals
		(a) FeSe crystal structure with only two constituent elements.
		(b) The structure in (a) reduces to paired quantum wells \((\alpha^{+}+\alpha^{-})\), whose localized electrons correspond to real-space Cooper pairs.
		(c) Assembled paired quantum wells form sublattice dichotomy, with upward/downward wells constructing a quasi-2D antiferromagnetic checkerboard lattice.
		(d) Tunneling current is governed by the 2-Fe unit cell holding two Cooper pairs, forming a localized bosonic system. Isolated \(2\alpha^{+}\) and \(2\alpha^{-}\) yield fermionic odd-parity \(dI/dV\) spectra (orange, blue dashed lines); their superposition gives the even-parity green solid curve. FeSe single crystals have a uniform well depth \(\xi_\alpha\), so their \(dI/dV\) traces show only one set of symmetric superconducting coherence peaks, consistent with the black experimental spectrum.}
	\label{figure4}
\end{figure}

As displayed in Fig. \ref{figure4}(a), single-crystal FeSe holds the simplest lattice among iron-based superconductors and hosts exotic unconventional properties \cite{hsu2008superconductivity}. Competing intertwined nematic, long-range antiferromagnetic and superconducting orders render it a vital platform to explore intrinsic iron-based superconducting mechanisms. Its ambient-pressure $T_\mathrm{c}\approx9\ \mathrm{K}$ can be raised to around $37\ \mathrm{K}$ under hydrostatic pressure \cite{Sun2016NatComm}. 
Within the quantum-well framework, bulk FeSe only stabilizes $\alpha$-type quantum wells with no observable $\beta$ phase. In Fig. \ref{figure4}(b), oppositely oriented $\alpha$ quantum wells form bonded pairs, where electrons form energy-degenerate real-space localized Cooper pairs. 
Fig. \ref{figure4}(c) presents the superconducting Fe plane, with interconnected quantum wells constructing a checkerboard bipartite sublattice. Arrows on each quantum well align with intrinsic long-range antiferromagnetic order. The red dashed square at the bottom-left corner indicates that such checkerboard geometry yields a two-Fe primitive cell, rather than the standard one-Fe crystallographic cell of pristine FeSe.

Based on the above analysis, we can qualitatively reproduce the STM spectrum of FeSe in Fig. \ref{figure4}(d). Since FeSe possesses a single quantum-well depth $\xi_\alpha$, its $\mathrm{d}I/\mathrm{d}V$ spectrum exhibits only one pair of superconducting coherence peaks at $\pm\Delta_\alpha$. As STM probes real-space local electronic states, we focus on four quantum wells confined within the 2-Fe primitive cell. The orange dashed curve with higher left edge and lower right edge denotes the electronic response of $\alpha^+$, while the blue dashed curve with lower left edge and higher right edge corresponds to $\alpha^-$. For bulk FeSe, the measured STM signal arises from superposition of contributions from stacked quantum-well layers, with substantial attenuation for deeper subsurface layers; this distorts the ideal U-shaped gap floor into the observed V-shaped profile in the central spectral region. The green solid line, obtained by summing the orange and blue dashed traces, accounts for Cooper-pair contributions from the two paired sets within one 2-Fe unit cell and perfectly matches the experimental black solid curve \cite{Kasahara2014PNAS}. Notably, individual $\alpha^+$ and $\alpha^-$ spectra feature left–right asymmetry with odd parity, whereas their superposition $(\alpha^+ + \alpha^-)$ restores symmetric lineshape with even parity after Cooper pairing.

\section{Odd Parity and Sublattice Dichotomy in Monolayer FeSe}

Epitaxial single-layer FeSe grown on SrTiO$_3$ (STO) substrates is the thinnest known iron-based superconductor with record-high $T_\text{c}$ among iron-based systems \cite{WangQY2012CPL}. Distinct from bulk single crystals in electronic properties, FeSe monolayers serve as a canonical two-dimensional platform for exploring unconventional high-$T_\text{c}$ superconductivity \cite{Tan2013,Lee2014}. The coupling of spin, orbital, lattice and charge degrees of freedom, together with the intrinsic superconducting mechanism, remains highly debated. Key open questions cover the origin of parity breaking and the microscopic pathway behind STO-induced drastic $T_\text{c}$ enhancement. Hu et al. put forward the sublattice dichotomy hypothesis \cite{Hu2013}, which only differentiates two inequivalent Fe sites and cannot fully capture dynamic interactions between superconducting electrons and external bias electric fields.
 Adopting a quantum-well picture, this chapter constructs an intuitive sublattice dichotomy model without distinguishing $\alpha$-Fe and $\beta$-Fe or performing tedious point-group analysis and Hamiltonian calculations. This model quantitatively reproduces all scanning tunneling microscopy (STM) spectral signatures and unveils the fundamental origin of odd parity. 
 
 \begin{figure}[tpb]
 	\centering
 	\includegraphics[width=\columnwidth]{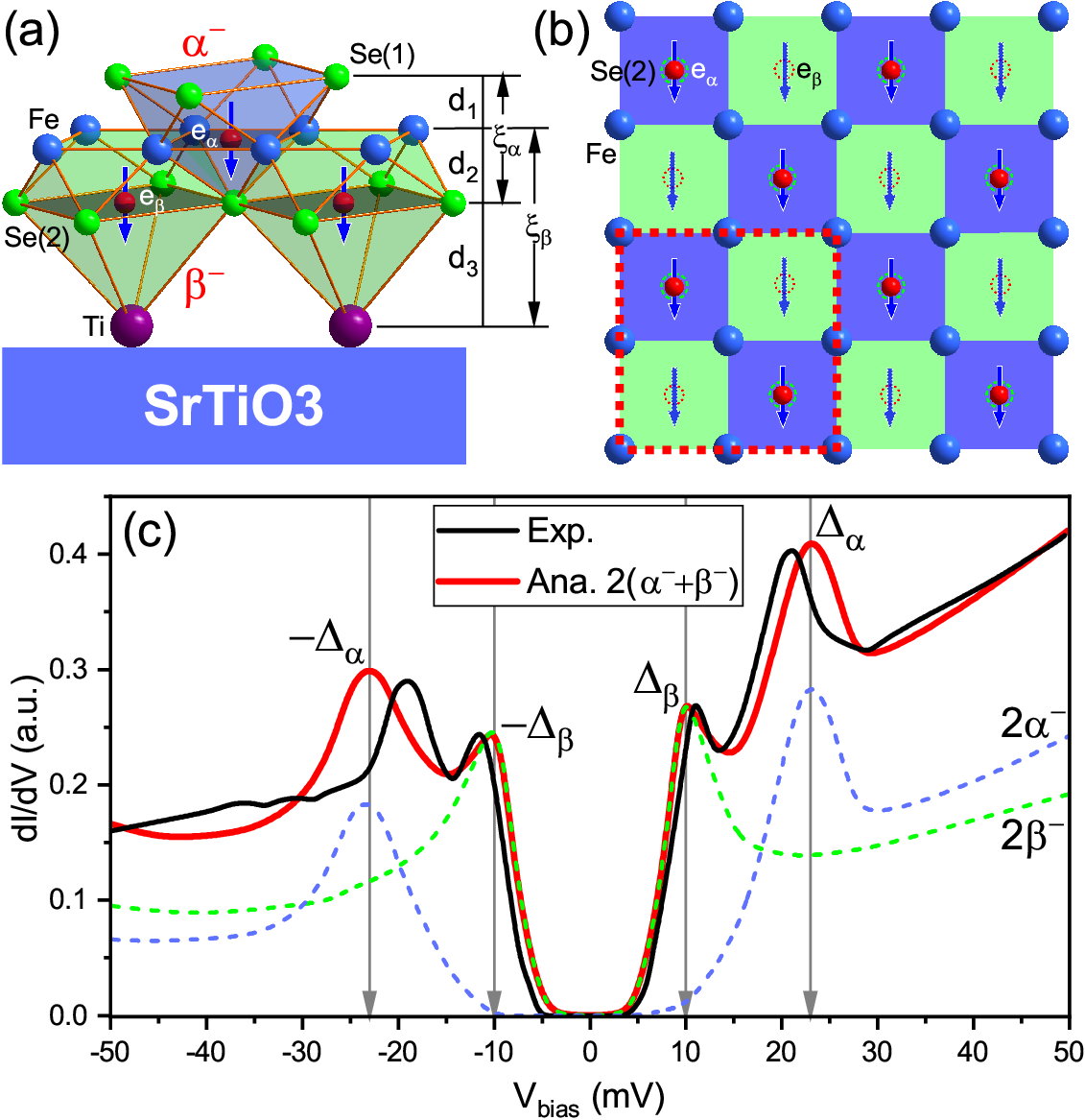}
 	\caption{
 		Quantum well sublattice dichotomy and odd-parity superconducting coherence peaks in monolayer FeSe films.  (a) Monolayer FeSe films couple to Ti bonds of the STO substrate, generating two downward quantum wells \(\alpha^-\) and \(\beta^-\) with depths \(\xi_\alpha\) and \(\xi_\beta\). This system supports dual superconducting phases and constitutes an odd-parity fermionic system.
 		(b) Self-assembled dual quantum wells produce sublattice dichotomy. Different from the Fe-layer antiferromagnetic order in Fig.~4(b), this bipartite checkerboard lattice possesses ferromagnetic interlayer coupling between Fe and Se(2) sublattices.
 		(c) Superconducting gap magnitudes \(\Delta_\alpha\) and \(\Delta_\beta\) are controlled by \(\xi_\alpha\) and \(\xi_\beta\). Analytical \(dI/dV\) spectra of the checkerboard unit cell (\(2\alpha^+ + 2\alpha^-\)) are shown as blue and green dashed lines. Their superposition yields the red solid line with asymmetric two-gap odd-parity superconducting interference peaks, which perfectly reproduces the black experimental curve.}
 	\label{figure5}
 \end{figure}

As demonstrated in Fig. \ref{figure5}(a), scanning transmission electron microscopy (STEM) characterizations confirm that the Se(1)/Fe/Se(2) monolayer bonds to the substrate primarily via a TiO-terminated interfacial layer. The substrate breaks the native mirror symmetry of the film. Assisted by Ti ions, two downward-oriented quantum wells $\alpha^-$ and $\beta^-$ form within the FeSe monolayer, with well depths $\xi_\alpha=d_1+d_2$ and $\xi_\beta=d_2+d_3$, respectively. Two distinct superconducting electronic states $e_\alpha$ and $e_\beta$ accordingly emerge, associated with two separate superconducting gaps and transition temperatures. As illustrated in Fig. \ref{figure5}(b), these two quantum wells form a checkerboard-like dual-sublattice configuration analogous to the single-crystal structure in Fig.~4(b), which avoids the need to postulate two inequivalent Fe atomic sites for physical interpretation.

Given experimental lattice parameters $d_1=1.43$ \AA, $d_2=1.35$ \AA, $d_3=2.92$ \AA \cite{Wang2020SciAdv}, the well depths are calculated as $\xi_\alpha=2.78$ and $\xi_\beta=4.27$, while experimental double superconducting gaps read $\Delta_1=21.3$ meV and $\Delta_2=11.1$ meV \cite{Kasahara2014PNAS}. Using the inverse-square relation in Eq.~(\ref{eq:delta}), we obtain two coefficients: $\eta_1=\Delta_1\xi_\alpha^2=164.6$, $\eta_2=\Delta_2\xi_\beta^2=202.3$. Their average serves as the undetermined coefficient $\eta(T)=(\eta_1+\eta_2)/2=183.5$. Substituting $\eta(T)$ back into Eq.~(\ref{eq:delta}) yields theoretical double gaps $\Delta_\alpha=23.1$ meV and $\Delta_\beta=10.1$ meV. 
Combining the quantum well orientations in Fig. \ref{figure5}(a), the $2\alpha^-+2\beta^-$ dual-sublattice unit cell in Fig. \ref{figure5}(b), and the U-shaped STM spectral profile of monolayer films, we extract the analytical $dI/dV$ curve of monolayer FeSe (red solid line) from data in Fig. \ref{figure3}(c), which agrees well with experimental spectra (black solid line). 
The core distinctions between bulk FeSe single crystals (Fig. \ref{figure4}) and monolayer FeSe films (Fig. \ref{figure5}) are summarized as follows:
(1) Bulk FeSe only hosts one set of degenerate paired quantum wells with identical geometry and opposite orientations, whereas monolayer FeSe contains two geometrically distinct quantum wells sharing the same downward orientation.
(2) Quantum wells in bulk FeSe preserve mirror symmetry, giving even-parity $dI/dV$ spectra. Broken mirror symmetry in monolayer FeSe triggers parity breaking and odd-parity spectral responses.
(3) Both systems form a checkerboard dual-sublattice lattice. Only Fe layers act as superconducting conductive layers in bulk FeSe, supporting one Bose-condensed superconducting state and one pair of coherence peaks. In monolayer FeSe, spatially separated Fe and Se layers both participate in superconductivity, generating two Fermi-condensed superconducting states and two pairs of coherence peaks.
Notably, the dual-sublattice character of both bulk and monolayer FeSe fundamentally originates from differences in quantum well size, population and orientation, without requiring an extra two-Fe-site assumption \cite{Hu2013}.

\section{Sublattice Trichotomy in 12442-Type Superconductors}

\begin{figure}[t]
	\centering
	\includegraphics[width=\columnwidth]{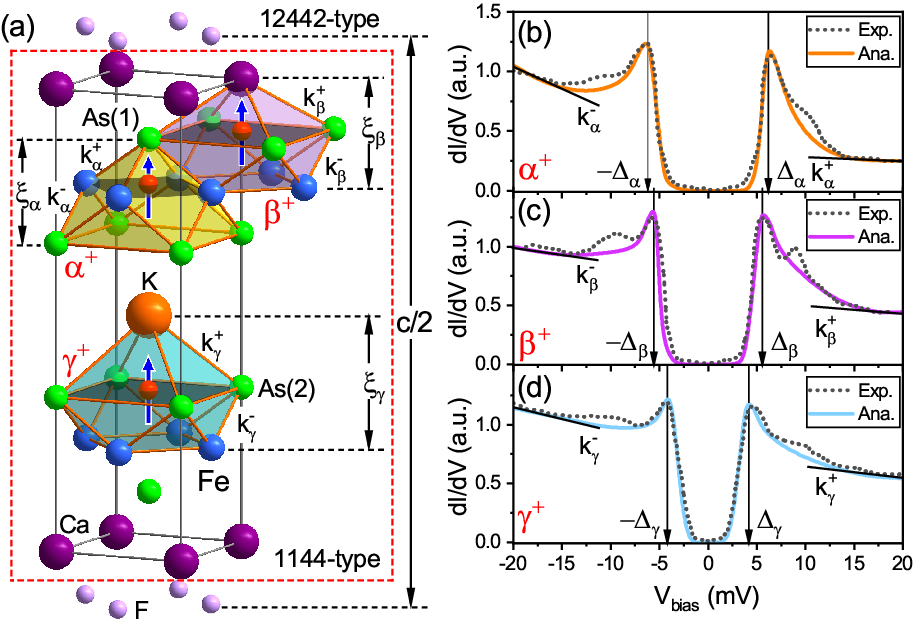}
	\caption{
		Triple quantum well sublattices and odd-parity STS spectra in 12442-type iron-based superconductors.
		(a) The \(\mathrm{KCa_2Fe_4As_4F_2}\) superconductor hosts three quantum wells with depths \(\xi_\alpha\), \(\xi_\beta\), \(\xi_\gamma\). Fe, As(1), As(2) layer-localized superconducting electrons generate three distinct superconducting states with individual gaps. Mirror symmetry breaking of each well is quantified by paired asymmetry parameters \((k^{+}_\alpha, k^{-}_\alpha)\), \((k^{+}_\beta, k^{-}_\beta)\), \((k^{+}_\gamma, k^{-}_\gamma)\).
		(b)–(d) Theoretical \(dI/dV\) spectra for \(\alpha^+\), \(\beta^+\), \(\gamma^+\) well states are contrasted with experimental data. All curves show asymmetric odd-parity line shapes with higher left spectral weight, a signature of upward-oriented quantum wells. Well depths follow \(\xi_\alpha < \xi_\beta < \xi_\gamma\), inversely correlated with superconducting gaps \(\Delta_\alpha > \Delta_\beta > \Delta_\gamma\).
		The \(\alpha^+\) spectrum in (b) features steeper slopes at large negative bias (\(k^{-}_\beta < k^{-}_\gamma < k^{-}_\alpha\)) and milder slopes at large positive bias (\(k^{+}_\alpha < k^{+}_\beta < k^{+}_\gamma\)) compared to \(\beta^+\) and \(\gamma^+\). This spectral trend solidly supports the quantum-well confinement picture and perfectly matches the black dashed experimental traces.}
	\label{figure6}
\end{figure}

Typical 1144 iron-based superconductors including $\text{CaKFe}_4\text{As}_4$ and $\text{EuRbFe}_4\text{As}_4$ \cite{meier2016caafe4as4} consist of double Fe–As slabs formed by staggered stacking of 122-type structural units. They host intrinsic superconductivity near 37 K without chemical doping \cite{singh2021bulk}. The observed three spin resonance peaks and odd-even electronic modulations offer key experimental support for the quantum-well superconducting picture \cite{xie2018odd}. 
Multiple derivative phases can be constructed based on the 1144 parent structure, and the 12442-type $\text{KCa}_2\text{Fe}_4\text{As}_4\text{F}_2$ is a representative example \cite{Wang2016JACS}. Its crystal structure is presented in Fig. \ref{figure6}(a). Insulating $\text{Ca}_2\text{F}_2$ fluorine slabs are inserted between the original 1144 structural motifs, separating neighboring FeAs layers. Despite this structural modification, the 12442 compound possesses the same quantum-well configuration as the 1144 parent, containing three independent quantum wells detailed as follows:
\begin{itemize}
\item $\alpha^+$ quantum well: As(1)/Fe/As(2) stacking, with Fe as the superconducting plane;
\item $\beta^+$ quantum well: Ca/As(1)/Fe stacking, where As(1) dominates superconductivity;
\item $\gamma^+$ quantum well: K/As(2)/Fe stacking, with As(2) hosting superconducting carriers.
\end{itemize}

Apparently, the $\beta^+$ and $\gamma^+$ quantum wells belong to the same type. The quantum-well depths satisfy $\xi_\alpha<\xi_\beta<\xi_\gamma$. Based on the quantum-well superconductivity theory, we predict that three distinct superconducting gaps satisfying $\Delta_\alpha>\Delta_\beta>\Delta_\gamma$ will definitely be observed in STS measurements of 12442-type iron-based superconductors. 
Recent STS experiments have indeed detected a three-gap feature in the $\text{KCa}_2\text{Fe}_4\text{As}_4\text{F}_2$ superconductor, with gap magnitudes $\Delta_1=6.2$ meV, $\Delta_2=5.4$ meV, and $\Delta_3=4.4$ meV \cite{Wang2016JACS}. Within our quantum-well framework, these gaps arise from three quantum wells of different characteristic depths: $\xi_\alpha=2.81$ \AA, $\xi_\beta=2.94$ \AA, and $\xi_\gamma=3.39$ \AA. The material parameter $\eta(T)$ is calculated from Eq. (\ref{eq:delta}) as $\eta_1 = \Delta_1 \xi_\alpha^2 = 48.9$, $\eta_2 = \Delta_2 \xi_\beta^2 = 46.7$ and $\eta_3 = \Delta_3 \xi_\gamma^2 = 50.5$. 
Taking the average of the three results gives the intrinsic material parameter $\eta(T) = (\eta_1+\eta_2+\eta_3)/3 = 48.7$. Using this averaged $\eta(T)$, we quantitatively reproduce the superconducting gaps of the three individual quantum wells via Eq. (\ref{eq:delta}): $\Delta_1=\Delta_\alpha=6.2$ meV, $\Delta_2=\Delta_\beta=5.6$ meV, and $\Delta_3=\Delta_\gamma=4.2$ meV. As presented in Figs. \ref{figure6}(b)--(d), our theoretical calculations agree excellently with STS experimental data.

Owing to the intrinsic mirror asymmetry of the quantum wells, electrons undergo completely different confinement effects during upward and downward transport. Such differences can be quantified by three groups of wavevector pairs $(k^{+}_\alpha, k^{-}_\alpha)$, $(k^{+}_\beta, k^{-}_\beta)$ and $(k^{+}_\gamma, k^{-}_\gamma)$. A larger $k$ indicates stronger electron confinement within quantum wells, smaller electron migration under applied bias, and weaker modulation of the polarization current. In differential conductance $dI/dV$ spectra, this feature corresponds to gentle signal evolution and shallow spectral slopes. 
By jointly considering the orientation, depth and mirror symmetry breaking of quantum wells, the slopes of high-bias $dI/dV$ curves obey definite ordering rules:
For positive bias, $k^{+}_\alpha < k^{+}_\beta < k^{+}_\gamma$;
for negative bias, $k^{-}_\beta < k^{-}_\gamma < k^{-}_\alpha$. Electrons under positive bias undergo strong direct vertex confinement, leading to markedly smaller $k$ values compared with those under negative bias. Figs. \ref{figure6}(b)–(d) compare the analytical $dI/dV$ signals from each quantum well with experimental data (black dashed lines). Three pairs of superconducting coherence peaks with antisymmetric odd parity are observed. Unlike the “low-left, high-right” profile in Fig. \ref{figure5}, the spectra in Fig. \ref{figure6} exhibit a “high-left, low-right” shape, which indicates opposite quantum-well orientations in the two material systems. In particular, the evolution of spectral slopes verifies the one-to-one correlation between the spectral features and the height as well as mirror asymmetry of quantum wells, further validating the reliability of the proposed new mechanism. In view of the lattice symmetry shown in Fig.~6(a), the 1144-type STM spectra can also present the right-low-left-high odd parity identical to that in Fig. \ref{figure5}. Furthermore, Cooper pairing between upward-oriented and downward-oriented quantum wells may produce even-parity superconducting coherence peaks with bilateral symmetry. These theoretical results need to be further verified by precise experimental measurements.

\section{Concluding Summary}

Departing from conventional quantum Hamiltonian frameworks, this work investigates the intrinsic correlation between superconducting scanning tunneling spectroscopy (STS) and crystal structures via symmetry breaking of the charge-neutral Mott-insulating ground state. Dissipationless supercurrent transport implies superconducting carriers behave as force-free energy insulators with five intrinsic properties: zero charge, zero rest mass, vanishing spatial size, temperature-independent invariant velocity, and arbitrary-barrier tunneling capability. Accordingly, the STM tunneling current arises from symmetry-induced electron–hole dipole polarization (essentially the applied bias field energy), rather than conventional real-space electron tunneling. 
STS features of iron-based superconductors are intrinsically dictated by quantum-well sublattices, contradicting the prevailing two-Fe sublattice dichotomy model for monolayer FeSe. Structural analysis of bulk FeSe, monolayer FeSe, and $\mathrm{KCa_2Fe_4As_4F_2}$ reveals one-, two-, and three-layer checkerboard quantum-well configurations, which precisely correspond to the experimentally observed one, two, and three pairs of superconducting coherence peaks. Mirror symmetry dictates spectral parity: bulk FeSe preserves full mirror symmetry, stabilizes real-space Cooper pairs and yields bosonic even-parity STS spectra. Interfacial symmetry breaking in monolayer FeSe suppresses Cooper pairing and produces asymmetric fermionic odd-parity spectral responses. 
The derived scaling relation $\Delta(T, \xi) = \eta(T)/\xi^2$ accurately reproduces the gap number and magnitude in diverse iron-based STS measurements. This work establishes a novel quantum-well physical picture and provides a fundamental framework for the unified theory of unconventional high-$T_\mathrm{c}$ superconductivity.

\noindent  \textbf{Data Availability Statement}: All relevant data is explicitly contained within the article.

\noindent \textbf{Competing interests}: The authors declare no competing interests.

\end{document}